\documentclass[12pt]{article}
\usepackage{amsmath}
\usepackage{amsfonts}
\usepackage{graphicx}
\usepackage{xcolor}
\graphicspath{ {../images/} }
\usepackage[margin=1in,top=1.5in,bottom=1.5in]{geometry}

\begin{document}

\title{Financial Markets and the \\
Phase Transition between Water and Steam}

\author{\\[0pt] Christof Schmidhuber\\ [20pt]
Zurich University of Applied Sciences\\ [0pt]
School of Engineering, Technikumstrasse 9\\ [0pt]
 CH-8401 Winterthur, Switzerland\\ [0pt]
christof@schmidhuber.ch\\  [30pt]}

\maketitle
\vspace{30pt}
{\bf Highlights}
\begin{itemize} \addtolength{\itemsep}{0 pt} 
\item The shares of financial assets are modeled by the molecules of a lattice gas
\item The lattice represents the invisible underlying social network of investors
\item In efficient markets, arbitrageurs drive the system  to its critical temperature
\item Quantum field theory is used to derive scaling laws in analogy with phase transitions
\item Consistency with empirical observations implies a fractal network dimension near 3
\end{itemize} 

\newpage
\begin{abstract}
A lattice gas model of financial markets is presented that has the potential to explain previous empirical observations of the interplay of trends and reversion in the markets, including a Langevin equation for the time evolution of trends. In this model, the shares of an asset correspond to gas molecules that are distributed across a hidden social network of investors. Neighbors in the network tend to align their positions due to herding behavior, corresponding to an attractive force of the gas molecules.\\  

This model is equivalent to the Ising model on this network, with the magnetization in the role of the deviation of the market price of an asset from its long-term value. Moreover, in efficient markets the system should naturally drive itself to its critical temperature, where it undergoes a second-order phase transition. There, it is characterized by long-range correlations and universal critical exponents, in analogy with the phase transition between water and steam. \\

Applying scalar field theory and the renormalization group, we show that these critical exponents imply predictions for the auto-correlations of financial market returns and for Hurst exponents. For a network topology of $R^D$, consistency with observation implies a fractal dimension of the network of $D\approx3$, and a correlation time of at least the length of the economic cycle. However, while this simplest model agrees well with market data on very long and on short time scales, it  does not explain the observed market trends on intermediate time horizons from one month to one year.\\

In a next step, the approach should therefore be extended to other models of critical dynamics, to general network topologies, and to the vicinity of the critical point. It allows us to indirectly measure universal properties of the hidden social network of investors from the empirically observable interplay of trends and reversion.
\\ \\

\end{abstract}

\thispagestyle{empty}
\newpage
\setcounter{page}{3}

\section{Introduction and Summary}

Trends in financial markets have long been exploited by the tactical trading industry \cite{turtles}, and they have also been well-studied in the literature (see, e.g., \cite{cutler,silber,fung,jaeger,miff,mosk,lemp,hurst,baz}). Traditional trend-followers are ``long'' a given market when its current trend is positive, and ``short'' when the trend is negative. Thus, they suffer draw-downs when trends revert.\\

In \cite{schmidhuber}, we have carefully measured the interplay of trends and reversion in financial markets with high precision by aggregating in new ways across 30 years of daily futures returns for equity indices, interest rates, currencies and commodities, and across trends over time horizons $T$ ranging from 2 days to 4 years. The key results are shown in fig. 1. A brief summary of how they were derived is given in the appendix.\\

We have found that trends tend to revert before they become statistically strongly significant. In other words, by the time a trend has become so obvious that everybody can see it in a price chart, it is already over. This is consistent with the hypothesis that any obvious market inefficiencies are quickly eliminated by investors. More precisely, we have defined the strength $\phi_T$ of a trend (\ref{phi}) measured over a time horizon $T$ as its statistical significance. We could accurately model tomorrow's return $R(t+1)$ in a market (normalized to have variance 1) by a quartic potential $V_T(\phi_T)$ of today's trend strength $\phi_T(t)$ in that market:\vspace{2pt}
\begin{equation}
R(t+1)\ =\ a\ -{\partial\over\partial\phi_T} V_T\big(\phi_T(t)\big)+\epsilon(t)\ \ \ \text{with}\ \ \ V_T(\phi_T)
=-{b_T\over2}\cdot \phi_T^2+{g_T\over24} \cdot \phi_T^4,\label{quartic}\vspace{5pt}
\end{equation}
\noindent
where $a$ is a risk premium, $\epsilon$ represents random noise, and $b_T$ and $ g_T$ are kinetic coefficients ($g_T$ is related to $c_T$ of \cite{schmidhuber} by $g_T=-6c_T$).
From this, we have derived a Langevin equation for the time evolution of the trend strength, which describes a damped motion in the potential $V_T(\phi)$ (see fig. 2).
We have interpreted $b_T$ as the persistence of trends, and $g_T$ as the strength of trend reversion. Within the limits of statistical significance, we have found the following (I thank J.-P. Bouchaud for pointing out that related observations were made in \cite{black}):
\begin{itemize}\addtolength{\itemsep}{-3 pt} 
\item
The parameters $b_T, g_T$ are universal in the sense that they are the same for all assets
\item
While $g_T$ has been fairly stable over time, $b_T$ appears to have vanished over the decades
\item
As a function of the time scale $T$, $g_T$ is approximately constant
\item
$b_T$ depends on the time scale: it peaks at $T\sim$ 3-12 months and decays for longer or shorter horizons. It appears to become negative for $T<1$ week or $T>$ several years.
\end{itemize}

\begin{figure}[h!]\centering
	\includegraphics[height=5.4cm]{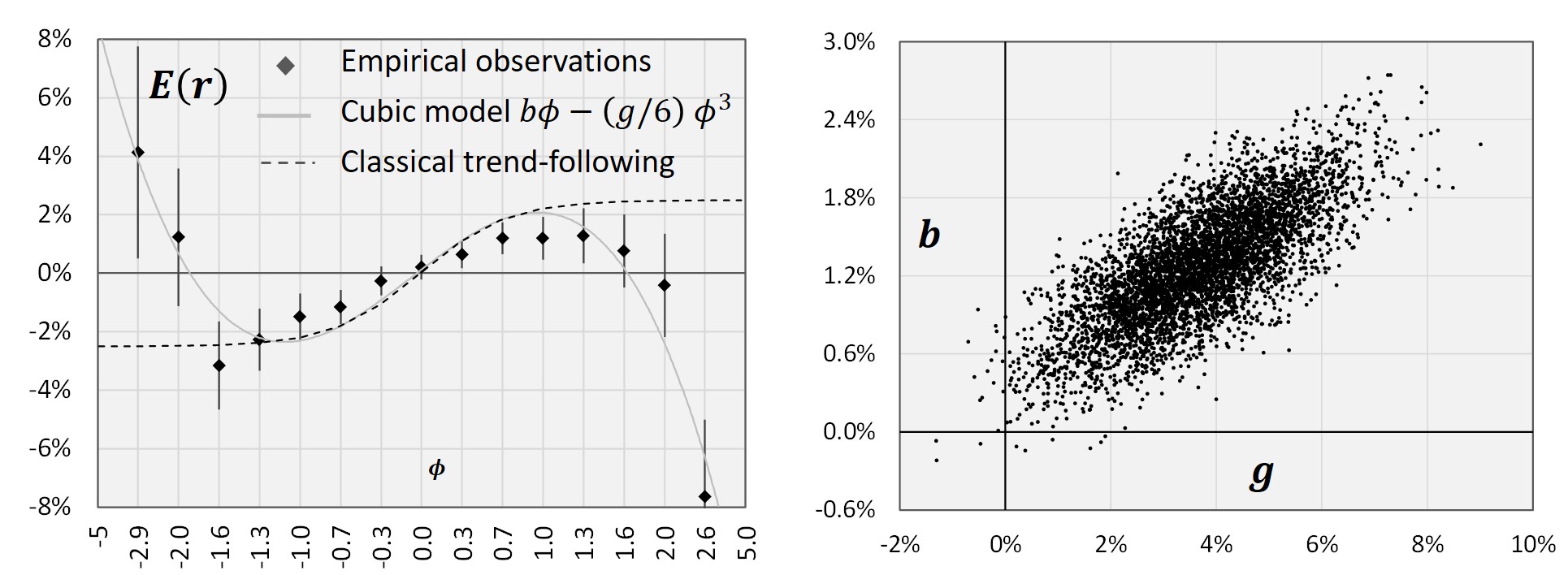}
	\includegraphics[height=5cm]{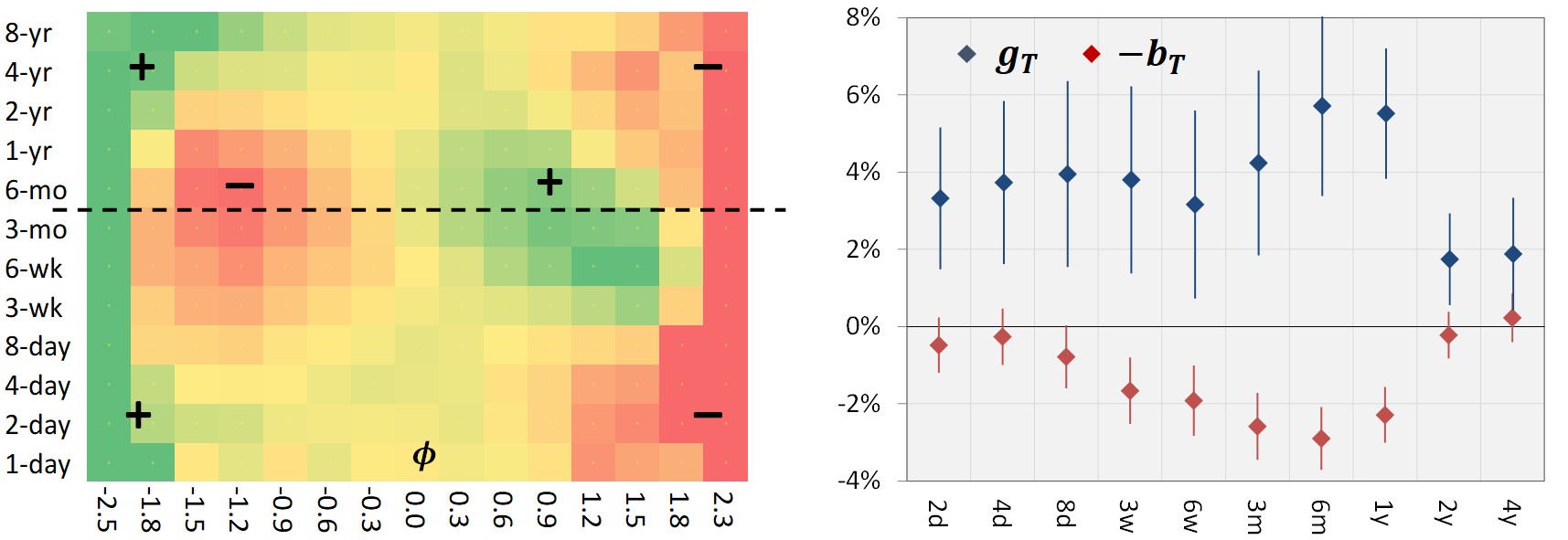} 
	\caption{
{\it Upper left (a)}: The expectation value $E(r)$ of the next day’s return of a futures market can be modeled by a cubic polynomial of the trend strength $\phi$, whose linear term $b\phi$ represents trend-persistence, and whose cubic term $-(g/6)\ \phi^3$ represents trend-reversion.  
{\it Upper right (b)}: As confirmed by bootstrapping, the regression coefficients $b$ and $g$ corresponding to the linear and cubic terms are statistically highly significant.
{\it Lower left (c)}: A heat map shows how the expectation value of tomorrow’s return depends both on today’s trend strength $\phi$ and its time horizon $T$. (a) can be thought of as a cross-section of (c) along the dashed line. 
{\it Lower right (d)}: The coefficients $b$ and $g$ as a function of the time horizon $T$, corresponding to the dashed line in (c) as it moves from 2 days to 4 years.}\label{figA}
\end{figure}

These results are reminiscent of physical systems at critical points. Stochastic processes with a quartic potential such as in (\ref{quartic}) are well-known to also describe the dynamics of the order parameter of many statistical mechanical systems near second-order phase transitions \cite{hohenberg}; see \cite{tauber} for a review. In this analogy, $V(\phi)$ is the Landau potential as a function of the order parameter. E.g., for a magnet, the order parameter is the overall magnetization. \\

\begin{figure}\centering
	\includegraphics[height=4.5cm]{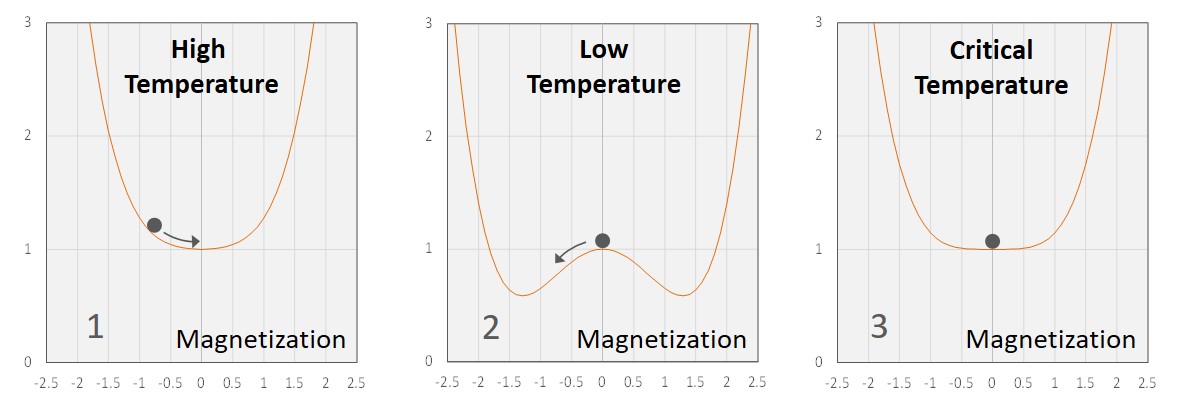}
\caption{
Landau potential of the Ising model as a function of the magnetization at high temperature (left), low temperature (center), and the critical temperature (right). }\label{figA}
\end{figure}

The classical such system is the Ising model \cite{ising}, which consists of magnetic spins on a square lattice. Spins can point either up or down, and neighboring spins tend to align (fig. 3, right). The effective potential $V(\phi)$ describes the energy of the system as a function of the magnetization $\phi$, i.e., of the average value of the spins (fig. 2). There is a low-temperature magnetized phase, where the potential has two minima corresponding to spins pointing either up or down, and a high-temperature disordered phase, where the average magnetization at the minimum of the potential is zero. The two phases are seperated by a second-order phase transition at some critical temperature. The time evolution of the magnetization follows a similar Langevin equation as we have derived for the trend strength $\phi$. \\

Of course, this could be a coincidence. On the other hand, analogies between financial markets and critical phenomena, such as scaling relations, have long been observed \cite{olson,glash,mant}. Much of the literature has focused on the scaling of volatilities rather than autocorrelations; see \cite{matteo} for a summary of empirical results and, e.g., \cite{drozdz} for a general review. 
 Such analogies are plausible, if financial markets are regarded as social networks, whose nodes correspond to idividual investors (fig. 3),
 who interact with each other in analogy with spins on a lattice, thereby creating the macroscopic phenomena of trends (herding behavior) and reversion (contrarian behavior). To replicate these phenomena, various spin- and agent models have been proposed (see, e.g., \cite{lux,farmer,bouchard}).\\
 
In fact, the idea of using the Ising model and its cousins to model social networks and financial markets has a long history (see \cite{sornette} for an extensive review). 
What is new about the results of \cite{schmidhuber} is that they make these long-standing ideas very specific, in the sense that we can now directly observe such a quartic Landau-like potential in financial markets, and we can empirically measure its coefficients and their scaling behavior. 
This allows us to make this conjectured relation between financial markets and phase transitions on a social network precise, and to test quantitatively, whether it agrees with financial market data. \\

\begin{figure}\centering
	\includegraphics[height=5.7cm]{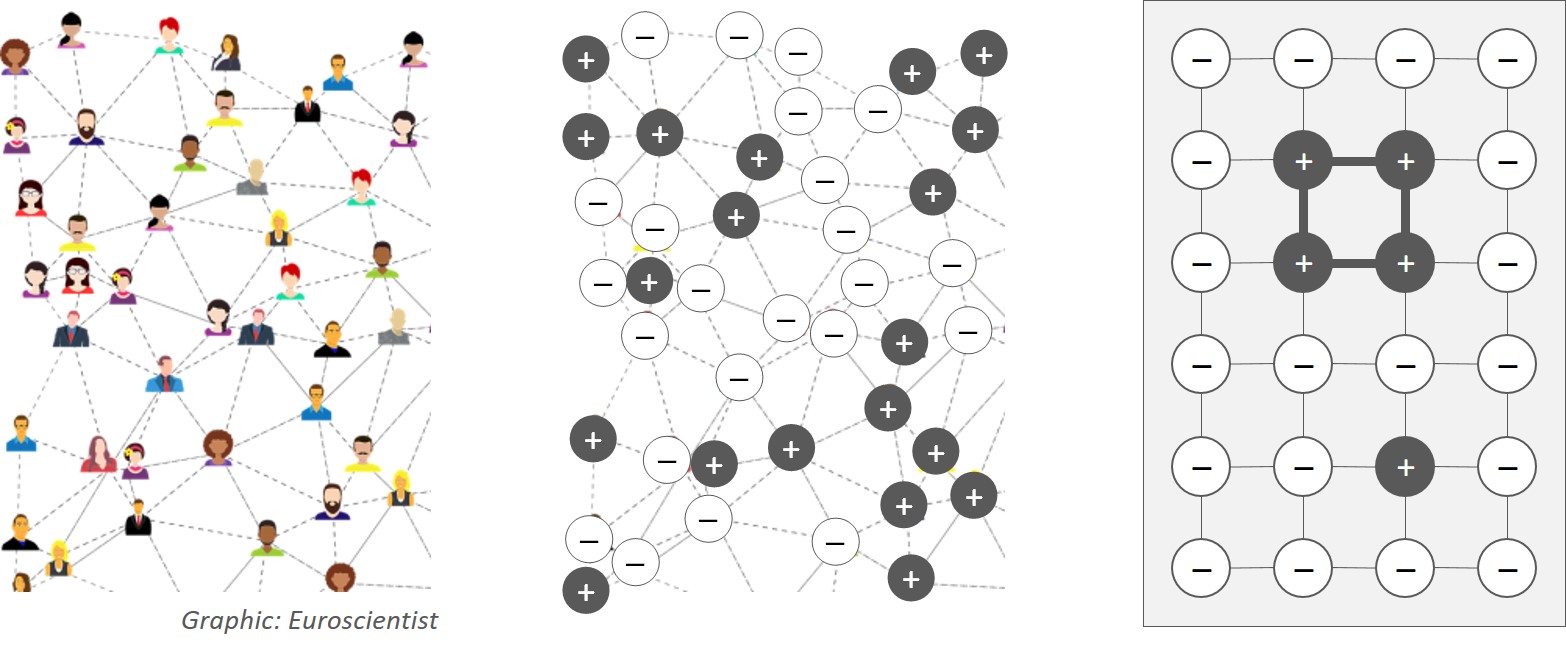}
\caption{Left: the social network of investors. Center: investors that are over/under-weight in a given asset relatively to the CAPM allocation are replaced by spins pointing up/down. Right: excerpt of an Ising model with 5 spins pointing up and energy $E=-4/2=-2$.}\label{figA}
\end{figure}

The first question is what exactly the microscopic degrees of freedom represent, which sit on the vertices of the network,
corresponding to individual investors. If we would identify up/down spins with buy/sell orders, 
as in the review \cite{sornette}, this would imply that the role of the order parameter (the magnetization) is played by the net order size, 
and not by the trend strength, as in our empirical observations.\\

We therefore take a different route. We identify the microscopic degrees of freedom with the relative portfolio weights, 
by which the investor sitting on a given node over- or underweighs a given asset. By ``relative'', we mean 
relative to the market capitalization of the asset, as in the CAPM model.
If we represent these relative weights by spins that can point either up (over-weight) or down (under-weight), and if we also assume that neigboring spins tend to align due to herding behavior, then we have all the ingredients that make a phase transition plausible, in analogy with the magnet. Moreover, we will argue that, in efficient markets \cite{mandel,samu,fama}, this system is driven to the critical temperature, at which the phase transition becomes second-order. It thus represents a form of self-organized criticality \cite{bak}.\\
 
We will show that this implies that the order parameter $\pi(t)$, integrated over the whole lattice (i.e., the overall magnetization),
corresponds to the deviation of the asset price from its long-term value. 
This deviation thus follows a Langevin equation with potential $V(\pi)$:
\begin{equation}
\dot\pi(t)\ =\ a\ -{\partial\over\partial\pi} V(\pi)+\epsilon(t)\ \ \ \text{with}\ \ \ V(\pi)
=-{b\over2}\cdot \pi^2+{g\over24} \cdot \pi^4.\label{price}
\end{equation}
We will further show how this can indeed be translated into an equation of the form  (\ref{quartic}) for the trend strength $\phi_T$, with coefficients $b_T, g_T$ that depend on the time scale.\\

Other models, in which the deviation of the asset price from its long-term mean moves in a polynomial potential 
have previously been considered \cite{ide,puk,puk1,halp2}. 
There, various potentials were found for shorter time windows, whose coefficients 
depend on the asset and the time window.  
By contrast, our quartic potential is measured over the whole 30-year time window, and is universal across assets. Indeed, if we assume that (\ref{price}) arises from the dimensional 
reduction of a higher-dimensional system near its critical point, this implies strong constraints on stochastic price processes  such  as (\ref{price})
and their scaling behavior:
it allows only processes that derive from renormalizable higher-dimensional field theories. \\
 
Thus, the network leaves its imprint on the asset price dynamics even after it is integrated over. 
This allows us to infer the properties of the presumed
unobservable social network that makes up financial markets from the observable interplay of trends and reversion. 
In the case where the network topology is assumed to be that of $R^D$, we find that 
the correlation time in financial markets must be at least of the same order of magnitude as the length of the economic cycle.
At shorter time horizons, there must be a ``scaling regime'' with a power-law fall-off of autocorrelations. Consistency with empirical observations,
such as Hurst exponents, then implies a (fractal) network dimension of approximately $D\approx3$.\\

However, upon closer look at the  auto-correlations of market returns, we also find that they cannot be fully explained by a simple Ising model on a network with the topology of $R^D$. In a next step, this approach should therefore be extended to other models of critical dynamics, as well as to more general network topologies.
Candidates for the latter include small-world networks \cite{watts}, scale-free networks \cite{barbasi}, or the Feynman diagrams of large-N field theory \cite{brezin}. For a recent review, see \cite{cimi}.
To our knowledge, no convincing specific model has emerged as a consensus so far. However, our results provide an empirical basis for accepting or rejecting such candidates: any statistical-mechanical model of financial markets, if accurate, must replicate the interplay of trends and reversion observed in \cite{schmidhuber}.\\

We hope that this will lead us to a specific and precise statistical model of the presumed social network of investors, thereby lead to a deeper understanding of financial markets, and - along the way - make the powerful tools of field theory accessible to finance.

\section{Lattice Gas Model of Financial Markets}

We begin with a rudimentary lattice gas model of financial markets. It is based on the following highly simplifying assumptions, to be refined later:
\begin{itemize}
\item The social network of investors is modeled by a hypercubic lattice of given dimension $D$. Each node corresponds to an investor.
\item There are only two assets in the world: (i) a bond, which can also be thought of as cash, and (ii) a single stock, whose shares represent the molecules of the lattice gas
\item Each investor can hold either one bond, or one share of the stock. The fact that an investor can hold at most one share represents a ``hard core'' of the gas molecules, or in other words, two molecules cannot sit at the same lattice site.
\item The long-term equity allocation of the average investor is 50\%, i.e., if there are $N$ lattice sites (investors), the average number of shares is $n_0=N/2$.
\item Neighbors tend to align their positions: if your ``friend'' ownes the stock, you tend to also be interested in it. This represents an attractive force between gas modelules.
\end{itemize}
A sample configuration of how shares may be distributed is shown in fig. 3 (right). Needless to say, the real network of investors is {\it not} a hypercubic lattice, and more generally, one may say that this is at best a rough cartoon of any real physical or financial system. However, it is well-known that this primitive lattice gas model can precisely describe real gases, such as water and steam or carbon dioxide, near their second-order phase transitions. There, all the microscopic details of these systems become irrelevant, resulting in a universal macroscopic behavior. Let us therefore assume that this phenomenon of universality also occurs on social networks. Later, we will allow for the possibility that these  universality classes must be refined for nontrivial network topologies. For now, let us quantify the above model.\\

For a given configuration, we define the occupation numbers $n_i\in\{0,1\}$ as the number of shares the investor $i$ holds. The total number of shares is $n=\sum_i n_i$. E.g., in fig. 3 (right), $n=5$. We also define the energy $E$ of a configuration $\vec n$ of occupation numbers as
\begin{equation}
E(n)=-{1\over D}\sum_{\langle ij\rangle} n_i n_j,\label{energy}
\end{equation}
where $\langle ij\rangle$ denote all pairs of neigbors, i.e., all links on the lattice. The normalization factor $1/D$ accounts for the fact that each node appears in $2D$ links. E.g., in fig. 3 (right), $E=-4/2=-2$, as there are four links with one share on both ends. 
This term represents the attractive force, as shares tend to lie next to each other to minimize the energy. 

We now imagine that the occupation numbers fluctuate while the total number of shares is kept fixed. In other words, shares can change owner (i.e., their location on the lattice). In addition, we assume that new shares can be issued or existing shares can be redeemed at a given cost $\mu$, which we set equal to 1. So we add a second term to the energy (\ref{energy}): 

\begin{equation}
E(n)=-{1\over D}\sum_{\langle ij\rangle} n_i n_j+\mu\sum_i n_i\ \ \ \text{with}\ \mu=1.\label{chemical}
\end{equation}
This term allows us to also let the total number of shares fluctuate, as in a system that is in equilibrium with a larger system. 
It is a counterweight to the attractive force (if $\mu$ was zero, the network would fill up with shares to minimize the energy). $\mu$ (or $-\mu$, depending on the convention) is called the ``chemical potential'' in statistical mechanics. \\

Where in this model is the time-dependent share price $P(t)$ that corresponds to real market prices? 
Since we have implicitly assumed that new shares are issued or redeemed at price 1 to satisfy any increase or decrease of demand, the share price is not dynamical. 
However, there is a dynamical market capitalization, namely the total number of shares $n(t)$, which is equivalent to a dynamical share price. \\

To make this equivalence precise, suppose that
the firm did not issue or repurchase shares. Thus, $n=n_0$ would be fixed, rather than fluctuate with the demand. Instead, there would be a share price $P(t)$ that dynamically adjusts to satisfy the demand. Then the market capitalization would be $n_0\cdot P(t)$. 
Putting both equations for the market capitalization together, we identify
\begin{equation}
P(t)={n(t)/n_0}\label{shareprice}
\end{equation}
as the dynamical share price, where $n_0=N/2$ is the long-term value of the firm.\\

As is well-known, the lattice gas model is equivalent to the Ising model. 
Replacing the occupation numbers $n_i$ by spins $s_i$, the energy (\ref{chemical}) simply becomes
\begin{equation}
E(n)=-{1\over D}\sum_{\langle ij\rangle} s_i s_j\ \ \ \text{with}\ s_i=n_i-1/2\in\{+1/2,-1/2\},\label{isingenergy}
\end{equation}
up to an irrelevant constant that depends only on the total number of lattice sites. The spins represent the over- or under-weight of a given investor in the stock (setting $\mu\neq1$ in (\ref{chemical}) would correspond to introducing an external magnetic field in the Ising model; here, we focus on the case $\mu=1$). From (\ref{shareprice},\ref{isingenergy}), the total magnetization
\begin{equation}
M(t)=\sum_i s_i = n(t)-{N\over2}={N\over2}\big(P(t)-1\big)\label{magnetization}
\end{equation}
of the Ising model is proportional to the deviation of the share price from its long-term value. \\

To summarize, we have re-interpreted the well-known Ising model in terms of financial markets. The order parameter (the overall magnetization) plays the role of the deviation of the share price from its long-term equilibrium value as defined, e.g., in the CAPM model. In the above, we have assumed a long-term market weight of 50\% for the asset; this can be generalized to an arbitrary market weight by assigning different probabilities $p$ and $(1-p)$ to the occupation numbers 0 and 1. \\

For $k$ assets, the spins have $k$ components.  E.g., for $k=3$, 
they can point in three different directions, corresponding to relative weights of a portfolio of three assets. It will be interesting to empirically measure whether there is an $O(k)$ symmetry for uncorrelated assets, in which case the Ising model would be replaced by a $O(k)$ vector model, such as a Heisenberg model for $k=3$.\\

As usual in statistical mechanics, we now consider the ensemble of all possible spin configuration $\vec s$, 
weighted with relative probability $G(\vec s )/Z$, where
\begin{equation}
G(\vec s)=\exp\{ -{1\over kT}E(\vec s)\}\ \ \ \text{and}\ \ \ Z=\sum_{\vec s} G(\vec s)\label{gibbs}
\end{equation}
where $G$ is the ``Gibbs factor'', and the ``partition function'' $Z$ acts as a normalization factor. The parameter $T$ (which is unrelated to the time horizon) controls how important the statistical fluctuations are. In physics, $T$ is interpreted as the temperature. $T$ is multiplied by Boltzmann's constant $k$ to give it the dimension of an energy. Below, we will relate $T$ to the persistence of trends $b$. Given (\ref{gibbs}), the expectation value of a spin $s_i$ is
$$\langle s_i\rangle = {1\over Z}\sum_{\vec s} s_i \exp\{ -{1\over kT}E(\vec s)\}.$$

\begin{figure}[t]\centering
	\includegraphics[height=8.5cm]{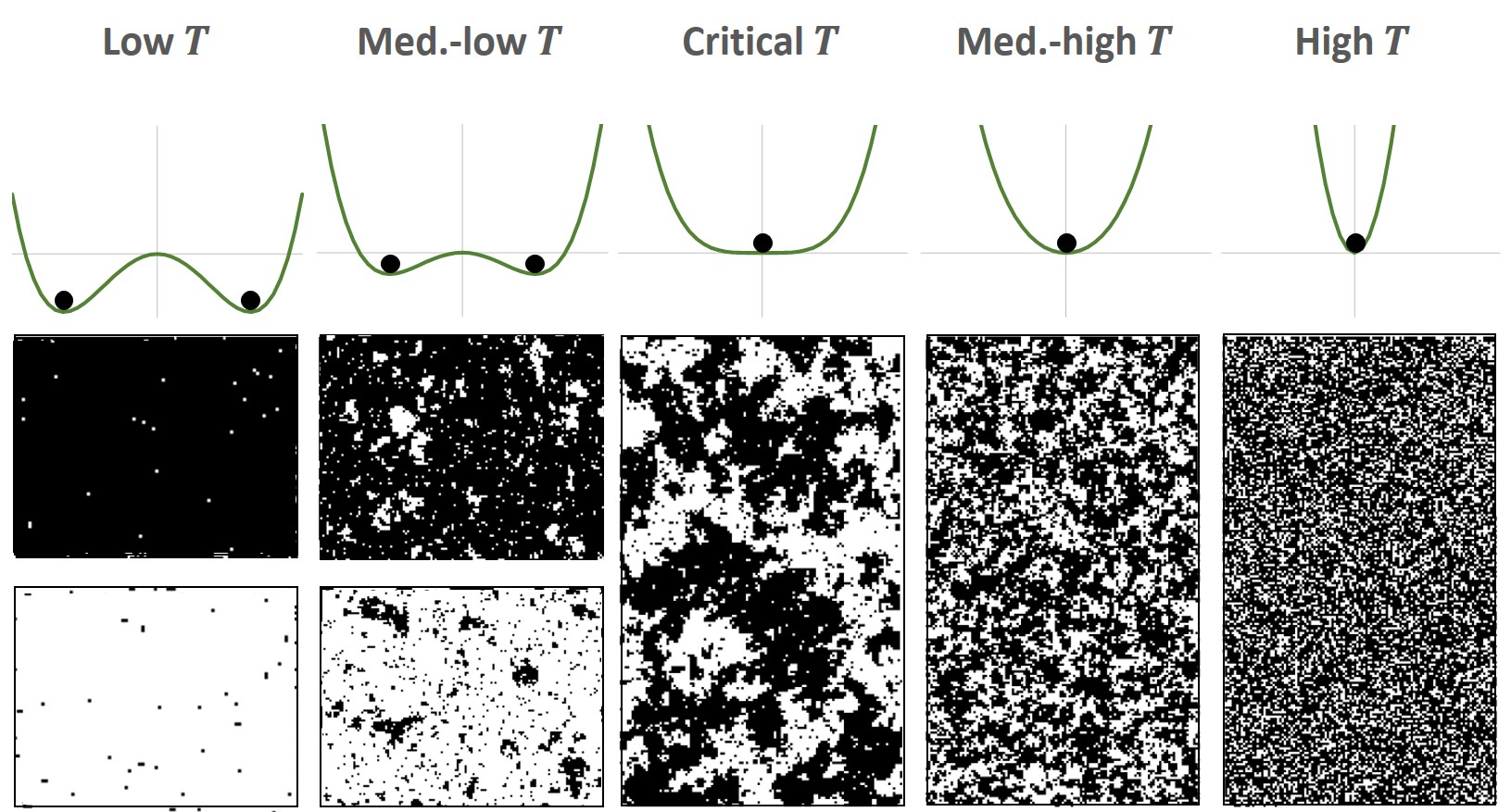}
\caption{Computer simulation of the lattice gas / Ising model. The upper graphs show the Landau potential. The lower graphs show typical spin configurations. If interpreted as assets in the network of investors, black dots represent shares, and white dots represent cash.}\label{figA}
\end{figure}

Fig. 4 shows images from a computer simulation of this model. Black dots represent shares, white dots represent cash. There is a high-temperature phase, corresponding to a strongly positive quadratic potential term, where all spin configurations are equally likely according to (\ref{gibbs}), and the spins are thus randomnly distributed. On the other hand, at low temperature, the energy (\ref{isingenergy}) is much more important than the statistical fluctuations in (\ref{gibbs}). So the lowest energy configurations dominate: all spins point either up or down, corresponding to the two minima of the potential with a strongly negative quadratic term. \\

As the temperature rises, statistical fluctuations become more important in (\ref{gibbs}) and bubbles - so-called Peierls droplets - start to form. 
Likewise, if the temperature is reduced starting from the high-temperature phase, spins start to feel the aligning force (\ref{isingenergy}) and clusters start to form.
Both the clusters and the bubbles have a typical size, the correlation length $\xi$. 
As we approach the critical point, where the quadratic term in the potential vanishes and the second-order phase transition occurs, the correlation length diverges: structures of all sizes can be seen, and the system is scale invariant. At and above the critical temperature, the
distinction between the two ground states disappears: one cannot distinguish whether one observes clusters of shares in a sea of cash, or clusters of cash in a sea of shares. \\

Let us now argue that financial markets are driven towards this critical point. 
First, suppose that the persistence of trends $b$, which corresponds to the quadratic term of the Landau potential, was significantly larger than zero, as in the low-temperature phase. 
Then trend-following would be so profitable that large amounts of capital would flow into this strategy. 
This would drive $b$ back to zero. 
On the other hand, if $b$ was significantly smaller than zero, markets would revert strongly und reliably.
As a result, mean reversion strategies (such as statistical arbitrage) would be so successful that a large amount of capital would flow into them.
This would again drive $b$ back to zero. \\
 
The important conclusion is that $b$ should be close to its critical value zero in efficient markets. 
In fact, a key empirical observation of \cite{schmidhuber} was that $b$, 
which was already small (of the order of a few percent) several decades ago, has further decreased since then. It now appears to be so close to zero
that even a diversified portfolio of pure trends can no longer be expected to yield significant excess return. 
This can indeed be explained by the fact that hundreds of billions, if not trillions of dollars, 
have flown into trend-following, thereby eliminating much of this market inefficiency.
It gives insight into how markets dynamically {\it become} efficient over time. The precise dynamical feedback mechanism that drives $b$ to 0 will be studied in seperate work. 
\\

Returning to the lattice gas, as we approach the critical temperature $T_c$ from below, the expectation value $\langle M(T)\rangle$ of the magnetization (\ref{magnetization}) goes to zero. 
If one measures how exactly it goes to zero, one finds a power law. Likewise, the correlation length $\xi$ diverges as a power:

\begin{equation}
M(T)\sim\vert \hat t\vert^\beta\ ,\ \ \ \xi(T)\sim \vert \hat t\vert^{-\nu}\ \ \ \text{with}\ \ \hat t\equiv(T-T_c)/T\label{betanu}
\end{equation}
with so-called ``critical coefficients'' $\beta,\nu$ that can be measured or computed. E.g., in $D=3$ dimensions one finds $\beta=0.33$ and $\nu=0.63$. 
The crucial property of $\beta,\nu$ is that they are ``universal''. E.g.,
our primitive lattice gas model in $D=3$ dimensions yields exactly the same values  $\beta,\nu$ as are observed for the second-order phase transitions of water and steam, of $CO_2$, 
and of all other systems in the same universality class. 
In fact, $\beta$ and $\nu$ depend only on 2 parameters: on the dimension $D$ of the lattice (but not on whether it is a square, triangular, or similar lattice), and on the symmetries. Here, the symmetry is $Z_2$, corresponding to flipping all spins. \\

A potential caveat is that this only applies to networks with the trivial topology of $D-$dimensional Euclidean space.
For more general network topologies, it is conceivable that the universality classes are characterized by additional parameters, 
such as the exponent $\gamma$ of the so-called ``degree distribution'' $p(k)\sim k^{-\gamma}$ of nodes with $k$ neighbors in the case of scale-free networks \cite{barbasi}. It will be important to investigate this in the future.\\

In either case, even though our lattice gas model of financial markets may seem very far from reality on a microscopic scale, the phenomenon of universality may allow it to precisely model the macroscopic behavior of real markets in the vicinity of the phase transition. 
In particular, the universal coefficients $\beta$ and $\nu$ in (\ref{betanu}) also determine how the system behaves under a rescaling in spare and time. 
 E.g., at the critical point, the covariance of spins at different points $\vec x,\vec y$ scales with their distance like
\begin{equation}
\langle s(\vec x) s(\vec y)\rangle\sim \vert \vec x-\vec y\vert^{2-D-\eta}\ \ \ \text{with}\ \ \eta=2{\beta\over\nu}+2-D.\label{eta}
 \end{equation}
 In the next section, we will see how $\eta$ can also be used to predict the auto-correlations of financial market returns at different points in time. Those can be compared with auto-correlations in real markets. This establishes a direct link between observable properties of asset prices, and the critical exponents of the presumed social network that makes up financial markets but cannot be observed directly.

\section{Field Theory of Financial Markets}

In the limit of infinitely many lattice sites with lattice spacing $a\rightarrow0$, the lattice becomes a continuous manifold of dimension $D$. 
Since networks generally have fractal dimensions, we formally allow $D$ to be any positive real number. 
The index $i$ becomes a $D$-dimensional vector $\vec x$, and the spins $s_i(t)$ become a field $\phi(\vec x,t)$ that measures the local magnetization, corresponding to the excess demand for an asset in a given neighborhood of the social network of investors. 
Let us first consider the system at a fixed point $t=0$ in time. One replaces the partition function $Z$ in (\ref{gibbs}) near the critical temperature $T\sim T_c$ by a path integral 
$$Z=\int D\phi(\vec x) \exp\{-S[\phi(\vec x)]\}$$
with action $S[\phi]$ and Landau potential $V(\phi)$:
\begin{eqnarray}
S[\phi(\vec x)]&=&\int d^D \vec x\  \Big\{{1\over2}\ (\vec\nabla\phi)^2+V(\phi)\Big\}\ \ \ \text{with} \ \ 
V(\phi)={1\over2} \ r \phi^2+{1\over24}\ g \phi^4.\label{action}
\end{eqnarray}
Here, the factor $1/kT$ in (\ref{gibbs}) has been absorbed such that $\phi$ has a standard kinetic term. The temperature then re-appears in the potential: $r$ is proportional to the deviation $(T-T_c)$ from the critical temperature. 
For a general introduction to field theory and critical phenomena, see, e.g., \cite{ZJ, kleinert}. Here, we merely list some results 
that we need in the following:
\begin{itemize}
\item The parameters $r(\lambda)$ and $g(\lambda)$ in (\ref{action}) ``flow'' under so-called renormalization group transformations \cite{wilson}, in which all distances are rescaled by a factor $\lambda$, and all objects with a scaling dimension are rescaled accordingly (see fig. 5).
\item In particular, for dimensions $D\ge4$, $g(\lambda)$ flows towards zero at large scales, so the quartic term of the potential becomes weaker and weaker. 
For $D>4$, the macroscopic scaling behavior of the system is the same as for a free theory (i.e., $g=0$). 
\item For dimensions $D<4$, $g(\lambda)$ flows to a nontrivial fixed point value $g^*$ at large scales (fig. 5). 
The coefficients of all other possible terms in the action, such as $\phi^6,(\vec\nabla\phi)^2$, ..., flow to zero. 
This explains why systems that are very different on a microscopic scale have the same macroscopic behavior with the same critical coefficients.
\item At this fixed point, the field $\phi$ scales as $\phi\rightarrow \lambda^{\eta/2}\phi$ with 
the so-called ``anomalous dimension'' $\eta=2(\beta/\nu)+2-D$.
$\eta$ can be computed perturbatively in $g^*$ via Feynman diagrams \cite{ZJ,kleinert}. Table 1 lists estimated values of $\eta$ for various dimensions $D$
\cite{ZJ,holovatch}.
\end{itemize}

\begin{figure}\centering
	\includegraphics[height=5.5cm]{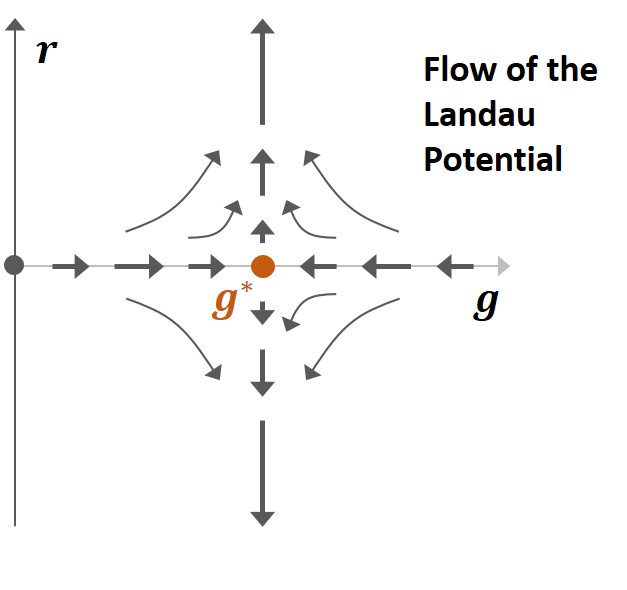}\hspace{30pt}
		\includegraphics[height=5.5cm]{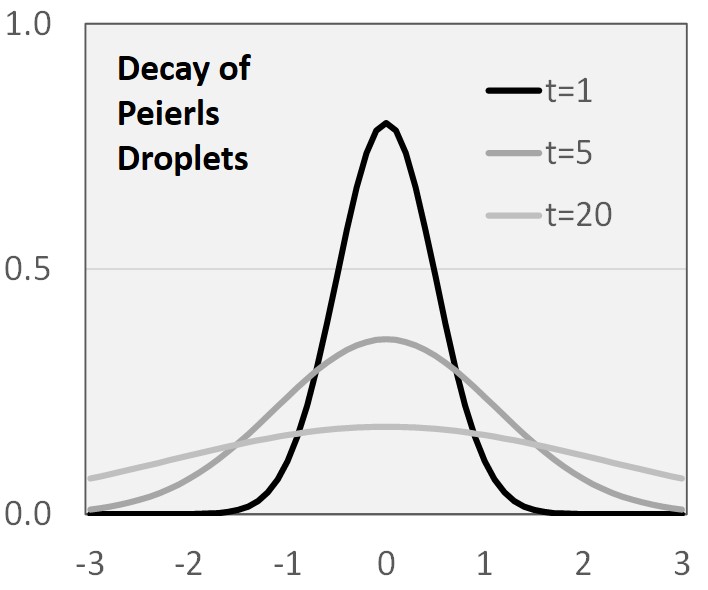}
\caption{Left: Renormalization Group Flow in $\phi^4$ theory. At large scales, the coefficient of the quartic term flows to a fixed point value $g^*$. The coefficient of the quadratic term must normally be fine-tuned to zero to reach the critical point. In our model, it is attracted to zero in efficient markets. Right: time evolution of Peierls droplets of positive magnetization.}\label{figA}
\end{figure}

Thus, the mere fact that we empirically observe a quartic potential in (\ref{quartic}) may already be a first indication that the fractal dimension of the social network of investors is less than 4.\\

\begin{tabular}{ |p{3cm}||p{1.5cm}|p{1.5cm}|p{1.5cm}|p{1.5cm}|p{1.5cm}|p{1.5cm}|	}\hline
	{\it Table 1 } &\multicolumn{6}{|c|}{Critical coefficients vs. network dimension $D$}\\	\hline
Dimension 	& 4  		&3.5 	& 3 		& 2.5	& 2 		 	& 1.5 	 	\\	\hline\hline
$\eta$			&	0.00	&	0.002	&	0.036	&	0.106	&	0.250	&		0.523		\\ \hline
$z$		   		&	2.000	&	2.001	&	2.024	&	2.071	&	2.167		&	2.352		\\ \hline
$\kappa\equiv(2-\eta)/z$	&	1.000	&	0.998	&	0.970	&	0.915	&	0.808		&	0.628		\\ \hline
\end{tabular}\\ \vspace{10pt}

The above points refer to static critical phenomena. Let us also discuss the time evolution of the system near the critical point, 
when it is slightly out of equilibrium. 
For a general introduction to this field of Critical Dynamics, see \cite{tauber}.   
In this paper, we focus on the simplest case - the so-called purely dissipative ``Model A'' \cite{hohenberg}. \\ 

By emerging and decaying, Peierls droplets of positive or negative magnetization create trends of the overall magnetization. 
In our model, where the magnetization represents the deviation of an asset price from its value, this is the origin of the price trends in financial markets.
The droplets represent groups of investors that collectively over- or underweigh an asset.  
The diffusion of the field $\phi(\vec x,t)$ in time is described by a Langevin equation \cite{ZJ,tauber}: 
\begin{equation}
{\partial\over\partial t}\phi(\vec x,t) ={\Omega\over2}\Big(\nabla^2\phi(\vec x,t)-r\phi(\vec x,t)-{1\over6}g \phi(\vec x,t)^3\Big)
+\text{random noise},\label{diffusion}
\end{equation}
where $\Omega$ is a diffusion coefficient and the variance of the noise is $\Omega$. By a rescaling of time, let us set $\Omega=1$.
\noindent
Fig. 5 (right) shows the diffusion of a droplet of size $L$ in the case of one dimension, corresponding to a local magnetization with Gaussian distribution. For $g=0$, it decays over a time span $\Delta t\sim L^2$. Correspondingly, the spatial correlation length $\xi$ and the correlation time $\tau$ are related by $\tau\sim\xi^2$. For nonzero $g$, there is a correction \cite{hohenberg}: 
$$ \Delta t\sim L^z\ \ ,\ \ \tau\sim\xi^z\ ,$$
where $z$ is a new critical exponent that depends on $D$ and the symmetry group $O(N)$. Its estimated value for various dimensions and $N=1$ (the Ising model) is shown in table 1. For a recent calulation, see \cite{hasenbusch}. For the range of fractal dimensions $D$ that will turn out to be relevant for us, $2\le D\le 3.5$, one may approximate $z= 2+c\cdot\eta$ \cite{hohenberg} with $c\approx2/3$.\\

In the following, we focus on the spatially constant mode of the field $\phi$, i.e., its integral over the whole network, corresponding to the overall magnetization:
$$\pi(t)=\int d^Dx\ \phi(x,t).$$
In the previous section, $\pi(t)$ has been identified with the deviation of the share price from its long-term value. 
Since this ``value'' is not observable, we cannot compare $\pi(t)$ directly with experiment. 
Fortunately, the unknown ``value'' drops out of the market returns $R(t)$, which are proportional to the price change $\dot\pi(t)$.
In the classical action (\ref{action}), $\phi$ is normalized to have standard kinetic term, so $\dot\pi$ has variance 1. On the other hand, the daily returns in our empirical analysis 
were also normalized to have variance 1. Thus, we directly identify
\begin{equation}
R(t)\equiv \dot\pi(t)={d\over dt}\pi(t) .\notag
\end{equation}
This immediately yields a first important prediction of $\phi^4$ theory for the auto-correlation of market returns that can be compared with observation:
\begin{equation}
\langle \dot\pi(t)\dot\pi(t')\rangle\sim\ -\ddot \Delta(t-t')
\ \ \text{with}\ \ \Delta(t)\equiv \langle\pi(0) \pi(t)\rangle, \label{autocorrelation}
\end{equation}
where $\Delta(t)$ is the so-called two-point function or ``propagator'' of the theory. 
Due to the huge amount of noise in financial markets, this autocorrelation of daily returns is very difficult to measure empirically.
For this reason, in our empirical observations \cite{schmidhuber} we have aggregated the recent returns of a given market over a time horizon $T=2/\omega$ in the form of the strength $\phi_\omega$ (\ref{phi}) of the recent trend in that market.
Its continuum definition in terms of $\dot\pi$ is:
\begin{eqnarray}
\phi_\omega(t)&=&2\omega^{3/2}\cdot\int_{-\infty}^{t-} d\zeta \ (t-\zeta)\ e^{-\omega (t-\zeta)}\ \dot\pi(\zeta)\label{theta}
\end{eqnarray}
Here, $t-$ indicates that the point $t$ is excluded from the integration, i.e., the return at time $t$ is predicted only from returns before time $t$. 
As explained in the appendix, the weight function $ne^{-\omega n}$, by which the return from $n\sim(t-\zeta)$ days ago is weighted in (\ref{theta}), peaks at $n=1/\omega=T/2$, and the average lookback period is $\langle n\rangle=T$. 
The pre-factor $2\omega^{3/2}$ ensures that the variance of the trend strength is approximately one, so $\phi_\omega$ measures the statistical significance (t-statistics) of the trend.
The aggregate auto-correlation can then be written as a (variant of the) Laplace transform of the propagator:
\begin{eqnarray}
\langle \phi_\omega(t),\dot\pi(t)\rangle=\langle \phi_\omega(0),\dot\pi(0)\rangle &=& 
-2\omega^{3/2}\int_{0+}^{\infty} d\zeta \ \zeta\cdot e^{-\omega \zeta}\ \ddot \Delta(\zeta).\label{auto}
\end{eqnarray}
Fig. 6 (left) shows the empirically observed two-point function $\langle \phi_\omega,\dot\pi\rangle$ on a logarithmic scale, with time horizon $T=2/\omega=2^k$ days.
It shows a curious behavior: at intra-month scales, the (very small) autocorrelation is negative. It turns positive on scales from one month to about 1 year - the time horizons, at which trend-following has been most profitable.  
At time scales of several years, it turns negative again. The correct field theory of financial markets must reproduce this propagator. 

\begin{figure}[h!]\centering
		\includegraphics[height=6cm]{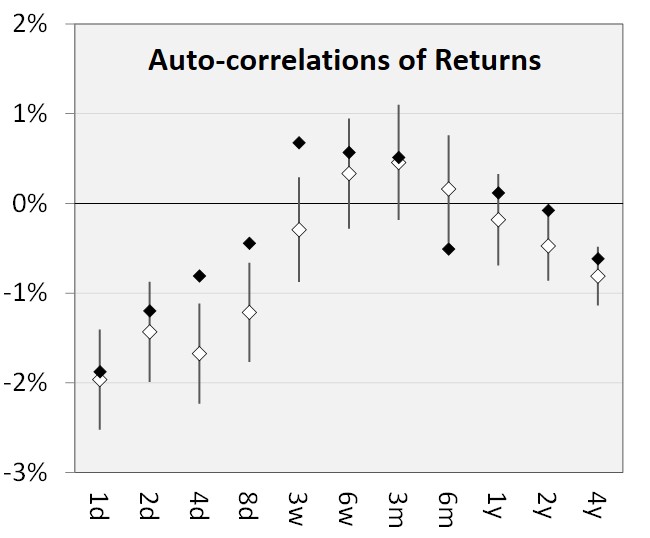}
				\includegraphics[height=6cm]{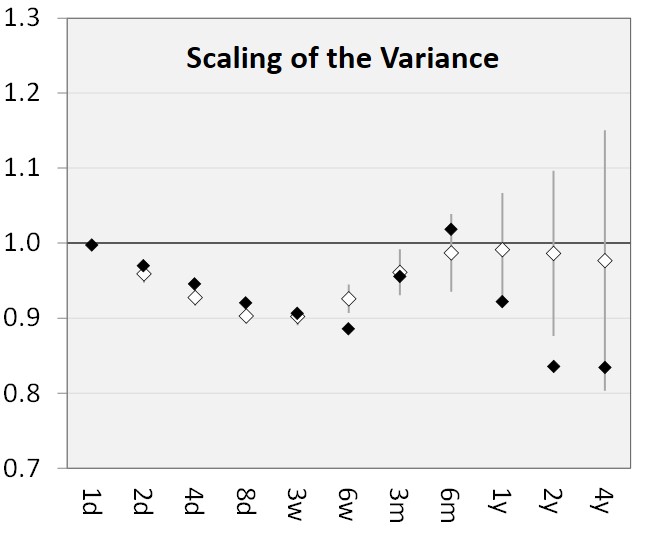}
\caption{ Left: autocorrelations of market returns (white dots), aggregated by the trend strength $\phi_\omega=\phi_{2/T}$. For intermediate time horizons, they turn positive, corresponding to the regime where trend-following works best. 
Right: the variance of the trend strength as a function of the time horizon falls off at short scales, rises at intermediate scales, and seems to fall off again at long scales. As a cross-check, similar analyses were performed 
with the simpler version of the trend strength $\tilde\phi$ (black dots) for non-overlapping time windows. }\label{figA}
\end{figure}

As an additional cross check, 
we have repeated the analysis with a simpler version $\tilde\phi_T$ of the trend strength (\ref{theta}), 
proportional to the equally weighted average of the daily log returns in a time window of length $T$
(this corresponds to the step function in fig. 9 of the appendix):
\begin{eqnarray}
\tilde \phi_T(t)&=& T^{-1/2}\int_{-\infty}^{t-} d\zeta \ \dot\pi(\zeta)\ =\ T^{-1/2}\big[   \pi(t)-\pi(t-T) \big].
\label{tilde}
\end{eqnarray}
To measure its auto-correlations empirically, we have slightly extended our history of futures returns from 30 years
to $2^{13}$ trading days $\approx31.5$ years, and we have constructed $2^{13-k}$ non-overlapping time windows of length $T=2^k$ for each $k$.
Fig. 6 (left) shows the correlation of this new trend strength (black dots) between neighboring time windows. Within estimation errors, these empirical results are in line with the empirical autocorrelations (\ref{auto}). The theoretical prediction for this correlation depends on the propagator:
$$\langle \tilde\phi_T(t)\tilde\phi_T(t-T)\rangle\ =\ -{1\over T}\big[\Delta(0)-2\Delta(T)+\Delta(2T) \big)].$$

Much of the literature on scaling in financial markets is concerned with the scaling of the variance, or more generally, of the $q$th moments of returns over the time period $T$ \cite{vicsek,mandel2}:
\begin{equation}
K_q(T)=\langle\vert \pi(T)-\pi(0)\vert^q\rangle/\langle\vert \pi(0)\vert^q\rangle\ \ \Rightarrow\ \ K_2(T)=2-2\Delta(T)/\Delta(0).\label{K}
\end{equation}
With a limited amount of data, these moments are easier to measure empirically than the auto-correlations of returns. Here we focus on $q=2$.  
From (\ref{autocorrelation}, \ref{theta}, \ref{tilde}), after some algebra we obtain formulas for the variance of our trend strengths:
\begin{eqnarray}
 \langle \phi_\omega(0)\phi_\omega(0)\rangle&=&
   -2\omega^3\int_{0+}^{\infty} du \ e^{-\omega u} \int_{0}^{u}dv\ v \dot\Delta(v).\label{normphi}
   \\  \langle \tilde\phi_T(0)\tilde\phi_T(0)\rangle&=&   {2\over T}\big[ \Delta(0)-\Delta(T)\big]  \ \ \ \text{with}\ \ T=2/\omega \label{normphitilde}
\end{eqnarray}
Note that these trend strengths were normalized to have variance 1, if the small auto-correlations of financial market returns are neglected. 
(\ref{normphi}, \ref{normphitilde}) thus describes the precise effect of the auto-correlations on this normalization, as predicted by field theory. \\

Fig. 6 (right) shows the empirical results. They mirror the curious behavior found for the propagator:
the variance of the trend strength as a function of the time scale decreases at short scales, increases at intermediate scales, then may decrease again.  
The correct field theory description of financial markets must reproduce this empirically observed variance.  \\

A further important test comes from the Langevin equation (\ref{diffusion}). For the spatially constant mode $\pi(t)$, which we have identified with the asset price minus its value, it implies:
\begin{equation}
\dot\pi(t)=-{r\over2}\cdot \pi(t)-{g\over12}\cdot \pi (t)^3+\eta(t), \label{lang1}
\end{equation}
where $\eta$ represents Gaussian noise. Again, $\pi$ is not directly observable because of the unknown ``value'' of an asset. 
However, this ``value'' cancels out of the trend strength. To compare with observation, we must translate (\ref{lang1})
into an equation involving only the trend strength. Generally, the coefficients will then depend on the time horizon $T=2/\omega$:
 \begin{eqnarray}
\dot\pi(t)&=&\beta_\omega\cdot \phi_\omega(t)+\gamma_\omega\cdot \phi_\omega(t)^3+\epsilon_\omega(t)\label{lang2}
\end{eqnarray}
The multiple regression coefficients $\beta_\omega,\gamma_\omega$ can be found by inverting the covariance matrix:
\vspace{5pt}\begin{eqnarray}
\begin{pmatrix} \beta\\ \gamma\end{pmatrix}
&=&\begin{pmatrix} \langle\phi^2\rangle&\langle\phi^4\rangle\\ \langle\phi^4\rangle&\langle\phi^6\rangle \end{pmatrix}^{-1}
\begin{pmatrix} \langle\phi,\dot\pi\rangle\\ \langle\phi^3\dot\pi\rangle \end{pmatrix}\notag\ =\ 
\begin{pmatrix} \langle\phi^2\rangle&\langle\phi^4\rangle\\ \langle\phi^4\rangle&\langle\phi^6\rangle \end{pmatrix}^{-1}
\begin{pmatrix} -{r\over2}\langle\phi,\pi\rangle-{g\over12} \langle\phi, \pi^3\rangle\\ 
-{r\over2}\langle\phi^3\pi\rangle-{g\over12} \langle\phi^3\pi^3\rangle \end{pmatrix},\notag
\end{eqnarray}
where all fields sit at $t=0$ (we have suppressed the index $\omega$). This reduces the problem to the (tedious but straightforward) computation of various correlation functions in $\phi^4$ theory, which is left for future work. \\

The correct field theory must replicate the empirically observed coefficients $b_T, g_T$ of (\ref{quartic}), as shown in fig. 1d. 
These results can be regarded as a refinement of the results for the propagator plotted in fig. 6 (left):
inserting (\ref{lang1}) into the expectation value $\langle\dot\pi(0)\dot\pi(t)\rangle$ decomposes the propagator into a component that comes from the persistence of trends $r$,
and another component that comes from the strength of reversion $g$. 
In particular, this should allow us to directly measure the fixed point value $g^*$, once we have found the right model.
As $g^*$ depends on the (possibly fractal) dimension $D$ of the presumed underlying social network of investors, this will amount to a direct empirical estimate of $D$.

\section{Illustration: $D$-dimensional Lattice}
 
In the previous section, we have derived three key requirements - yielding the correct propagator, the correct variance of the trend strength, and the correct Langevin equation - that a field theory of financial markets must satisfy in order to agree with observation.\\

In this section, we illustrate how to apply these criteria in practise at the simplest example, to be generalized in future work: 
the purely dissipative model A on a hypercubic lattice as discussed in section 3. We do {\it not} necessarily expect this model to closely match observation. 
One of the reasons is that the social network of investors is surely much more connected than a hypercubic lattice, where the shortest path between two sites is on average very large. 
However, this example demonstrates how to connect the rather abstract concepts of quantum field theory with the practical observation of financial markets in order to derive two key properties of the presumed underlying social network of investors: 
its fractal dimension $D$, and its correlation length $\xi$. To this end, we discuss two distinct regimes: 
\begin{itemize}
\item
the ``scaling regime'' at small time scales $T$ (i.e., $\omega=2/T\rightarrow\infty$). Here, 
the renormalization group dictates a power-law fall-off of all correlation functions. Matching with observation should yield an estimate of the anomalous dimension $\eta$, and thereby of $D$. 
\item
the ``exponential regime'' at large time scales (i.e., $\omega\rightarrow0$). Here, all correlations fall off exponentially. Matching with observation should yield an estimate of the correlation time $\tau$ of market prices,
and thereby of the correlation length $\xi\sim\tau^{-z}$ in the network.
\end{itemize}
These two regimes also appear in more general models of critical dynamics, albeit with different scaling exponents. 
The precise interpolation between the two regimes at intermediate time scales should hold the key to identifying the correct model, beyond ``model A'' on a hypercubic lattice. 
This is left for future work. In this paper, we only use the fact that the scaling regime must approach the exponential regime, as the dimension approaches the critical dimension $D=4$.\\

Sub-section 4.1 discusses the propagator of the model. 
Sub-section 4.2 turns to the scaling of variances and compares the theoretical predictions with the observed variances and Hurst exponents. 
Sub-section 4.3 discusses the results.

\subsection{The Propagator} 

We consider ``model A'' on a Euclidean manifold of dimension $D$ (representing the network of investors) and set the variance $\Omega$ of the random noise to 1. For the free theory ($g=0$, corresponding to $D>4$), the propagator of the spatially constant mode $\pi(t)$ is \cite{tauber}:
$$G_{cl}(p,t)={1\over p^2+r}\ \exp\{-{p^2+r\over2}\vert t\vert\}\ \ \Rightarrow\ \ \Delta(t)={1\over2}\cdot{\xi^2} \exp\{-\vert t\vert/\xi^2\},$$
where $\xi$ is the correlation length, which is equal to $\sqrt{2/r}$ for the free theory with $g=0$.
For $g\neq0$ and $D<4$, the renormalization group provides the following scaling relation for the exact two-point function in the critical domain, slightly above the critical temperature (where $\xi$ in (\ref{betanu}) is finite) \cite{tauber}:
$$\Delta(t)=\langle\pi(0)\pi(t)\rangle = \xi^{2-\eta}\cdot g(t/\xi^z),$$
with some function $g(x)$ that falls off exponentially at large $x$.  Let us first consider the scaling regime. 
Consistency with the static limit $t\rightarrow0$ (where $\Delta(0)\sim\xi^{2-\eta}$ from the spatial integral over $\vec x$), and with  the limit $D\rightarrow4$ (where $\kappa\rightarrow1$ as in table 1) implies:
\begin{equation}
\Delta(t) ={1\over2}\cdot\tau^\kappa-{1\over2}\cdot \vert t\vert^\kappa\ \ \ \text{for } t\ll\tau\ \ \ \text{with} \ \ \tau\equiv\xi^z,\ \ \kappa\equiv{2-\eta\over z}<1\label{propagator} 
\end{equation}
On the other hand, in the exponential regime, consistency with the limit $D\rightarrow4$ implies
\begin{equation}
\Delta(t) ={1\over2}\cdot \tau^\kappa \exp\{-\vert t\vert/\tau\}\ \ \ \text{for } t\gg\tau\label{propagator1} 
\end{equation}

\begin{figure}[t]\centering
		\includegraphics[height=7cm]{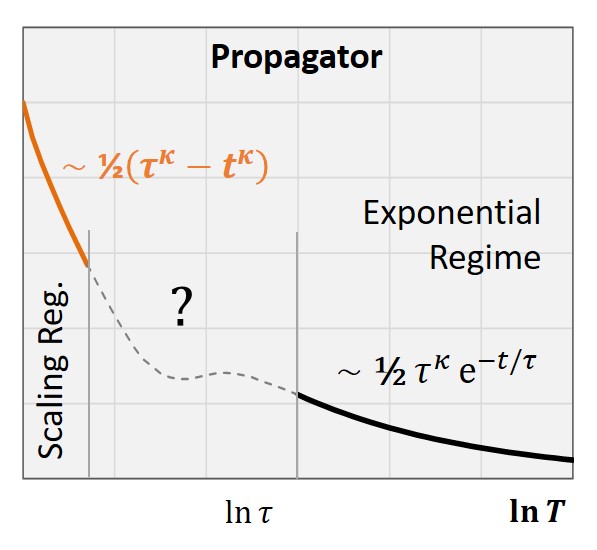}
\caption{Left: the propagator of the spatially constant mode falls off like a power at small time scales, and exponentially at time scales larger than the correlation time. 
The unknown behavior of the correct model in the intermediate regime is indicated by a question mark. 
}\label{figA}
\end{figure}

In the low-temperature phase, these relations hold for the connected two-point function $\langle \pi(0)\pi(t)\rangle-\langle \pi(0)\rangle^2$ instead. 
In both cases, we have for $t>0$:
\begin{eqnarray}
\dot\Delta(t) &=&\left\{\begin{array}{ll}
        -{1\over2}\tau^{\kappa-1} \cdot\exp\{- t/\tau\}  \\
        -{1\over2}\kappa\cdot t^{\kappa-1}\\
\end{array}\right. ,\ 
\ddot\Delta(t) =\left\{\begin{array}{ll}
        {1\over2}\tau^{\kappa-2}\cdot\exp\{- t/\tau\} & \text{for } t\gg\tau\\
        {1\over2}\kappa(1-\kappa)\cdot t^{\kappa-2} &\text{for } t\ll\tau\\
\end{array}\right. \label{derivatives} 
\end{eqnarray}
Note that this holds both above and below the critical temperature: 
the propagator in these cases differs only by the expectation value of $\pi$, which is a constant that drops out of $\ddot\Delta$.\\

Fig. 7 schematically shows this propagator. The singularity of the orange line is the imprint that the invisible higher-dimensional social network of investors leaves on the stochastic process $\pi(t)$,
even after it is integrated out. As $D\rightarrow 4$ and $\kappa\rightarrow1$, the propagator in the scaling regime approaches that of the exponential regime, and the singularity disappears.\\

\noindent
The propagator (\ref{propagator}, \ref{propagator1}) predicts that auto-correlations $-\ddot\Delta$ (\ref{autocorrelation}) of market returns are purely negative. Thus, it does not explain the positive autocorrelations 
that we observe in fig. 6 (left) on intermediate scales from one month to one year. 
These empirical results suggest that the true propagator
$\Delta$ has negative curvature in part of the intermediate regime, which creates the phenomenon of market trends.
Again, this should be a key clue for finding the true model in the future.  

\subsection{Scaling of the Variance}

For the time being, though, let us nevertheless proceed and discuss the scaling regime and the exponential regime,
where our model is in line with observations. To this end, it is more convenient to consider the scaling 
of the  variances (\ref{normphi}, \ref{normphitilde}) of the trend strengths instead of the propagator. 
Fig. 8 (left) extends the empirically estimated variance of fig. 6 to 16-year windows.
From (\ref{normphi}, \ref{normphitilde}, \ref{derivatives}), we obtain in the scaling regime $T=2^k={2/\omega}\ll\tau$ for $\kappa\approx1$:
\begin{eqnarray}
{\langle \phi_\omega^2\rangle}&\sim&\omega^{1-\kappa}=e^{(1-\kappa)\cdot\ln\omega} \ \approx\ 1-(1-\kappa)\ln 2\cdot k\label{varshort}\\
\langle\tilde\phi_T^2\rangle&=&T^{\kappa-1}= e^{(\kappa-1)\cdot\ln T}\approx 1-(1-\kappa)\ln 2\cdot k\label{vartilde}
\end{eqnarray}
Thus, $\kappa$ can be read off from the slope $(1-\kappa)\ln 2$ of the variance as a function of $k$ (orange line in fig. 7). We measure by regression:
\begin{equation}
\kappa=0.96\pm0.01\ \ \ \Rightarrow\ \ \ D=2.9\pm0.1.\label{dimension}
\end{equation}
Here, we have interpolated the formally non-integer dimension $D$ from table 1 based on the approximation $z=2+\eta(D)\cdot 2/3$.
We propose to identify $D$ with the fractal dimension of the network; although there are different definitions of this fractal dimension, they coincide for homogenous networks such as ours, where all nodes have the same number of links (for other network topologies, this is generally not the case, though \cite{song}). \\

\begin{figure}[t]\centering
		\includegraphics[height=6cm]{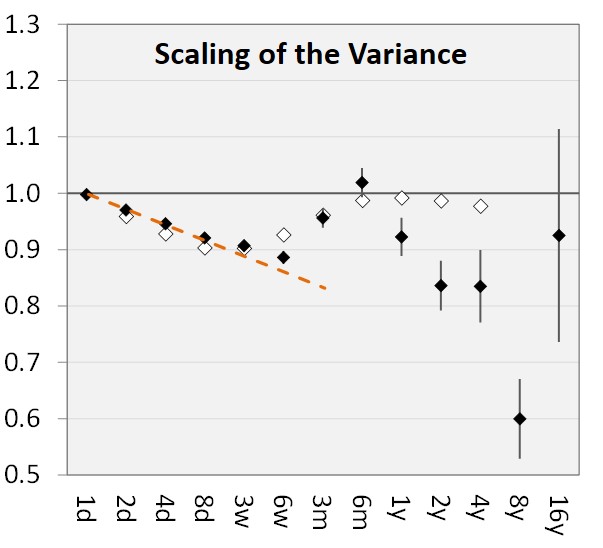}
		\includegraphics[height=6cm]{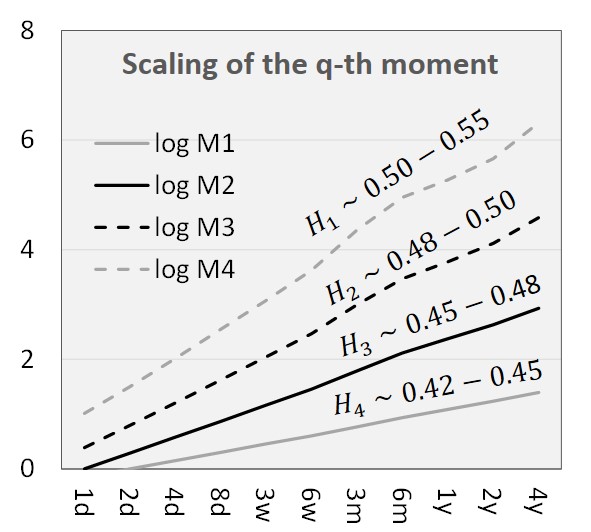}
\caption{Left: Fit of the scaling behavior of the variance in the scaling regime (dashed orange line) by the field theory prediction with critical coefficient 
$\kappa\approx0.96$, corresponding to a fractal dimension of the presumed underlying network of $D\approx3$ . The correlation time,
where the variance falls off rapidly, must be at least as long as the economic cycle. 
Right:  Scaling of the $q$-th moment $M_q$ for $q=\{1,2,3,4\}$ as a function of the time scale $T$ in our dataset of 
liquid daily futures returns. The slopes of the lines are the Hurst coefficients $H_q$. 
}\label{figA}
\end{figure}

On the other hand, in the exponential regime $T={2/\omega}\gg\tau$ we get from (\ref{normphi}, \ref{normphitilde}, \ref{derivatives}):
\begin{eqnarray}
{\langle \phi_\omega^2\rangle}\  =\  { \omega^2\over(\omega+1/\tau)^2}\cdot \tau^{\kappa-1}\ \ \ ,\ \ \ 
{\langle \tilde\phi_T^2\rangle}\  =\  {\tau\over T}(1-e^{-T/\tau})\cdot \tau^{\kappa-1} \label{varlong}
\end{eqnarray}
In both cases, field theory predicts the variance of trends to fall off rapidly as $T$ becomes larger than the correlation time. There is some indication from fig. 8 (left) that this 
happens at $T\sim 8$ years, but we do not have enough data to confirm this with significance, 
and we cannot even rule out that the correlation time is infinite. 
All we can say is that
the correlation time must be larger than $4-8$ years. Interestingly, this lower bound is of the same order of magnitude as 
the length of the economic cycle. 
A more precise statement will require an analysis of many decades, if not centuries of historical data.

Let us also comment on Hurst exponents.  
We use the definition of the $q$-th generalized Hurst exponent $H_q$ in terms of
the scaling of the $q$-th moment of $\pi$ (see \ref{K}):
$$M_q(T)\equiv\langle\vert\pi(T)-\pi(0)\vert^q\rangle\sim \langle\vert\pi(0)\vert^q\rangle\cdot T^{qH_q}\ \ \ \text{with}\ \ q\in R,$$
The process $\pi$ is called ``mono-scaling'', if all $H_q$ are identical, and ``multi-scaling'', if $H_q$ depends on $q$.
There is ample evidence that financial market returns are ``multi-scaling'', not ``mono-scaling'' (see \cite{matteo} and references therein, and \cite{drozdz} for a general review of scaling properties, including in financial markets). \\

Our data set also supports this. 
Fig. 8 (right) shows the empirical measurement of $M_q(T)$ for $T$ ranging from 1 day to 4 years
for our 30-year history of daily futures returns across asset classes. 
From the slopes of the lines, we can estimate the generalized Hurst exponents. The results are consistent with those for liquid markets in \cite{matteo} (our dataset only covers liquid futures). Our Hurst exponents are not quite the same for all horizons, though. E.g., $H_2$ increases from 0.48 at short scales to 0.5 at scales of several years.\\ 

What does field theory predict for the generalized Hurst exponents? We focus on even, integer $q=2n$. 
In the scaling regime, we get from (\ref{tilde}, \ref{vartilde}): 
$$M_{2n}(T)\ =\ T^{n}\langle\vert\tilde\phi_T^2\vert^n\rangle\ \sim\ T^{n\kappa}\ \ \ \text{with}\ \ \ n\in\{1,2,3,...\}$$
Thus, for infinite correlation time, field theory predicts mono-scaling with Hurst exponent
\begin{equation}
H={\kappa\over2}\ \ \ \text{with}\ \ \ \kappa\ =\ {2-\eta\over z}\ \label{hursteta}
\end{equation}
independently of $n$. Thus, the Hurst exponent is directly related to the anomalous scaling dimension $\eta$ of the deviation $\pi$ of the asset price from its value, which in turn is related to the dimension $D$ of the underlying network of investors.
E.g., for $D=(2,3,4)$, field theory predicts $H\approx(0.40, 0.485, 0.50)$.
However, any finite correlation time will destroy this mono-scaling. How precisely the $H_{2n}$ depend on $n$ in this case, and whether this can explain the observed multi-scaling in financial markets, must be addressed in future work.

\subsection{Discussion}

To summarize this section, we have tried out the simplest model of critical dynamics (``model A'') on the simplest network topology (a $D$-dimensional hypercubic lattice) at the critical point (where $g=g^*$ is at its fixed point) in order to model financial markets
as a social network of investors. Although these assumptions were 
radical simplifications that seem unrealistic, the predictions of this model are not all that far from the reality of financial markets.\\

At small scales, it predicts a power-law fall-off of autocorrelations 
and a scaling behavior of volatilities that is reminscent of the one that has been observed in financial markets. 
In our approach, the origin of the associated scaling laws and of Hurst exponents that differ from 1/2 is the same as in second-order phase transitions:
they reflect anomalous dimensions that are well-known from quantum field theory. We have demonstrated how this allows us to infer the
(fractal) dimension of the presumed underlying social network of investors.
It is encouraging that the estimated fractal dimension $D\approx3$ is in line with reported fractal dimensions of other social networks, the majority of which lie between 2 and 4 \cite{song}.\\

Moreover, it is highly nontrivial that the small typical values of the anomalous dimension $\eta$ have just the right magnitude to the explain the observed small negative 1-day market auto-correlations of a few percent. 
We take this as a sign that we are on the right track by modeling financial markets in analogy with critical phenomena. At the same time, we have seen that ``model A'' does not describe the intermediate regime at horizons from a few weeks to a few years well.
We may need a more complex model, or a more complex network topology, or we may see non-universal properties of the network (see the concluding remarks).\\

At large scales, our model predicts an exponential fall-off of all correlations. This should be verified by studying historical market data on a scale of decades and centuries. Of course, the question is then whether markets in previous centuries can already have been efficient enough to 
be driven to the critical point by the mechanism described in section 2, and whether the structure and dimension of the network of investors has been stable in time.

\section{Conclusion and Outlook}

Motivated by the previous empirical observations \cite{schmidhuber} on trends and reversion in financial markets, in this paper we have proposed a statistical-mechanical model of the markets that has the potential to explain these observations. 
The key clue is that financial market returns can accurately be described by the stochastic process (\ref{quartic}) with a quartic potential.   
Such a quartic potential typically occurs in certain physical models near second-order phase transitions, such as the Ising model or the lattice gas model. \\

This has led us to propose a lattice gas model of financial markets. The lattice represents the social network of investors, who correspond to its vertices. 
The microscopic degrees of freedom sitting at these vertices represent the investors' asset allocations.
They can either be modeled by gas molecules that correspond to the shares of an asset, 
or - equivalently - by spins that denote whether the investor is over- or underweight in the asset, relatively to its CAPM weight. For $N$ different assets,
the spins have $N$ components.\\

We have identified the order parameter, corresponding to the magnetization in the spin model, as the deviation of the asset price from its long-term value.
If we assume that neighboring spins tend to align due to herding behavior, the interaction of the spins naturally creates the macroscopic phenomena 
of trends and reversion of the asset price.\\

In addition, we have argued that efficient markets drive themselves to the critical point, at which a second-order phase transition occurs. 
This assumes that investors quickly exploit and thereby eliminate any obvious market inefficiencies 
either by momentum strategies, such as trend-following, or by reversion strategies, such as statistical arbitrage. This argument is supported by the observation that 
the persistence of trends, corresponding to the quadratic term in the quartic potential, has gradually disappeared over the decades \cite{schmidhuber}.\\

There is a wide variety of stochastic processes such as (\ref{quartic}) that have been studied in finance \cite{shreve}. 
Different assets are modeled by different processes, which are often complex and require the calibration of many parameters to fit the historical data. 
What is currently missing is a unified theory that determines from first principles, which stochastic process with which parameters should be used to model a given asset. 
Our approach suggests such a principle: these processes must arise from the dimensional reduction of a higher-dimensional system near its critical point,
at which it is described by a renormalizable (self-consistent) quantum field theory.
This key criterium of renormalizability leaves us with only a small set of universal field theories, and therefore of allowed stochastic processes. \\

More precisely, at this critical point, the system is characterized by a divergent correlation length 
and by critical coefficients, which are universal in the sense that - in the case of a trivial network topology - they depend only on two parameters: the fractal dimension $D$ of the network, and the symmetry group $O(N)$.
We have used the methodology of field theory and the renormalization group to relate these critical exponents to observable autocorrelations and Hurst exponents of financial market returns. In particular, the second Hurst exponent follows from the anomalous scaling dimension $\eta$ and the dynamic critical exponent $z$ (\ref{hursteta}):
$$H_2={\kappa\over2}\ \ \ \text{with}\ \ \ \kappa= {2-\eta\over z}$$
This allows us to measure $\kappa$, and from its value infer the fractal dimension of the 
presumed underlying social network of investors. \\

As an illustrative example, we have applied this procedure to the simplest case: ``Model A'' on a network with the topology of $D$-dimensional Euclidean space.
Consistency with observation then yields the following ``predictions'' for the fractal dimension $D$ of the network and for the correlation time $\tau$ in financial markets (\ref{dimension}):
$$D\approx 3\ \ ,\ \ \tau \ge\ \text{length of the economic cycle}.$$
This can in principle explain the scaling observed in financial markets at short and very long time scales, and the small magnitude of market auto-correlations.
Encouragingly, $D$ is within the range of fractal dimensions typically found in social networks \cite{song}. \\

On the other hand, by comparing the two-point function of this model with the observed auto-correlations of financial market returns at intermediate time scales, 
we have also seen that the universal properties of model A on a network with Euclidean topology do not explain the trends observed in real markets
over time horizons from 1 month to 1 year. This leaves us with the task of refining the model, such that it can explain the observed interplay of trends and reversion also on intermediate time scales. Some natural refinements are:
\begin{enumerate}
\item {\it Beyond ``Model A''}: one may try to model the asset allocations of investors by other degrees of freedom than spins, e.g., by a binary fluid in the case of two assets, leading to different universality classes.
More generally, the models ``A-J'' of critical dynamics and their generalizations provide a rich laboratory of systems \cite{hohenberg,tauber}.
The universal properties of some of them indeed include two-point functions that switch between positive and negative auto-correlations, as observed in fig. 6 (left). 
\item {\it Nontrivial network topologies}: we do expect the social network of investors to be much more connected than a hypercubic lattice, 
for which the average distance between two points is very long. Candidates for social networks include scale-free networks \cite{barbasi},
which are characterized by a ``degree distribution'' $p(k)\sim k^{-\gamma}$ of nodes with $k$ neighbors. 
It is currently not clear, which of these networks can be part of renormalizable field theories (except for the planar diagrams of large-N field theory \cite{brezin}, on which the renormalization group flow has been studied \cite{me}). An important next step is thus to classify their universality classes. 
Apart from the dimension $D$ and the symmetry group $O(N)$, they may also depend on parameters such as $\gamma$. 
\item {\it Approach to the critical point}: it is also conceivable that the presumed underlying social network of investors is not exactly at the fixed point $g=g^*$.  In this case, we may be able to understand the oscillations of the two-point function in fig. 6 (left) as non-universal phenomena that arise during the {\it approach} to the critical point. 
\end{enumerate}

In all three cases, it will be fascinating to search for the right candidate in analogy with the example of section 4, by using
the different models of critical dynamics to predict the scaling behavior of market returns, and by then 
ruling out those models that are inconsistent with empirical observation. Further work in this direction is underway.

\section*{Acknowledgements} 

I would like to thank Uwe Täuber for very helpful remarks and explanations on dynamic scaling, and Wolfgang Breymann for encouragement, valuable discussions, and for pointing out the relation to self-organized criticality. I also thank Didier Sornette for an inspiring conversation and Michel Dacorogna for comments on the manuscript. This research is supported by grant no. CRSK-2 190659 from the Swiss National Science Foundation. \\

\newpage
\section{Appendix: Brief Review of Empirical Results}

In \cite{schmidhuber}, we have empirically observed the interplay of trends and reversion in financial markets.
The analysis is based on historical daily log-returns of a set of 24 futures contracts \cite{mendel}. 
This set is diversified across four asset classes (equity indices, interest rates, currencies, commodities), three regions (Americas, Europe, Asia) and three commodity sectors (energy, metals, agriculture). 
We used futures returns, instead of the underlying market returns, because futures returns are guaranteed to be marked-to-market daily. 
This is a brief suammary of the approach and results. For detailed explanations and references, we refer to \cite{schmidhuber}.

\subsection{Trend Definition}

For each of the 24 markets $i$, we consider 30 years of daily price data $P_i(t)$ from 1990 to 2019. We define normalized daily log-returns $R_i(t)$: 
\begin{equation}
R_i(t)={r_i(t)\over \sigma_i}\ ,\ \ \ r_i(t)=\ln {P_i(t)\over P_i(t-1)}\ ,\ \ \ \sigma_i^2=\text{var}(r_i)\ ,\ \ \ \mu_i=\text{mean}(r_i)\ ,\label{return}
\end{equation}
where the long-term daily risk premium $\mu_i$ and the long-term daily standard deviation $\sigma_i$ of a market $i$ are measured over the whole 30-year period.
For each market and on each day, we compute 10 different trend strengths on 10 different time scales $T_k$ with
\begin{equation}
T_k=2^k\ \text{business days with}\  k\in \{1,2,3,...,10\}\notag
\end{equation}
This represents time horizons of approximately 2 days to 4 years. At a given point in time, a market may well be in an up-trend on one scale, but on a down-trend on another scale.  
For a given horizon $T$, we define the trend strength $\phi_{i,T}(t)$ of a given market $i$ at the close of trading of day $t\in Z$ as a weighted average of past daily log-returns of that market (i.e., on or before day $t$), in excess of the long-term risk premium
in that market:
\begin{equation}
\phi_{i,T}(t)=\sum_{n=0}^\infty w_T(n)\cdot \hat R_i(t-n)\ \ \ \text{with}\ \ \ \hat R_i(t-n)=R_i(t-n)-{\mu_i\over\sigma_i},\notag
\end{equation}
where $w_T(n)$ is a weight function for the time scale $T$ (to avoid any biases, in the out-of-sample cross-validation of our results, the risk premia were estimated only from the training samples, excluding the validation samples).

The weight function $w_T(n)$ is always normalized such that the trend strength $\phi_{i,T}$ has standard deviation 1. Assuming that market returns on different days are independent from each other (which is true to high accuracy), this implies: 
\begin{equation}
\sum_{n=0}^\infty w_T^2(n)=1.\notag
\end{equation}
With this normalization, $\phi_{i,T}$ can be regarded as the statistical significance of the trend. E.g., $\phi_{i,T}=2$ represents a highly significant up-trend, while $\phi_{i,T}=-0.5$ represents a weakly significant down-trend. This normalization makes all trend strengths comparable with each other, and thus allows us to aggregate across different markets and time scales.\\

The simplest weight function would be a {\it step function} (fig. 1, dotted line). In this case, the trend strength $\phi_{i,T}$ is just proportional to the average log-return over the past $T$ days. A wedge-like weight function (fig. 1, solid line)
would correspond to a {\it moving log-price average crossover}. 
The advantages and disadvantages of different weight functions are disussed in \cite{schmidhuber}. 
For our study, we used two such definitions of the trend strength $\psi_T$ and $\phi_T$: 
\begin{eqnarray}
\psi_{T}:&&\tilde w_T(n)=M_T\ e^{-2n/T} \ \ \ \text{with normalization} \ \ M_T=\sqrt{1-e^{-4/T}},\label{psi}\\
\phi_{T}:&&w_T(n)= N_T\cdot (n+1)\cdot \exp(-{2n\over T})\ \ \ \text{with}\ \ \ N_T={(1-e^{-4/T})^2\over\sqrt{1-e^{-8/T}}}.\label{xex}\label{phi}
\end{eqnarray}
The "average lookback period" of these trend strengths, i.e., the expectation value $E[n+1]$ of the number of days we look back (where "today", i.e. $n=0$, counts as a 1-day lookback), is
$T/2$ for $\psi$ and $T$ for $\phi$.
Actually, for a given horizon $T$, the precise definition of the trend strengths does not matter much - they all yield quite similar results in our regression analysis, as long as the weight function rises gradually, decays gradually, and the average lookback period is the same.
However, (\ref{psi},\ref{phi}) were best-suited for our study, because they
\begin{itemize} \addtolength{\itemsep}{-5 pt} 
\item have only a single free parameter, the horizon $T$ (to avoid overfitting historical data)
\item can be computed recursively (which yields a Langevin equation for continuous time)
\end{itemize}
(\ref{phi}) also has the advantage of reducing trading costs. It was originally
introduced by the author in 2008 at Syndex Capital Management, and has been used to
replicate Managed Futures indices as part of a UCITS fund from 2010-2014.

\begin{figure}[h]\centering
	\includegraphics[height=4.5cm]{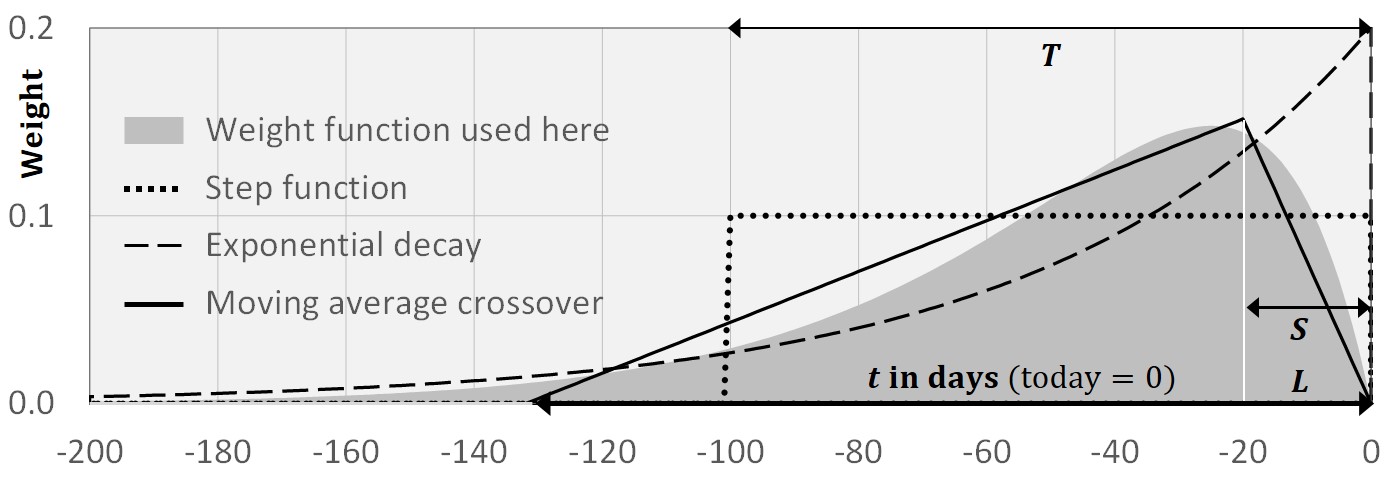} 
	\caption{Our trend strength is defined as a weighted sum of past log-returns. The grey area shows the weight function used in this paper, compared with three standard alternatives. All four weight functions shown here have the same average lookback period.}
	\label{figA}
\end{figure}

\subsection{Prediction of next-day Return}

The results shown in fig. 1 are aggregated across 30 years of daily returns for each market, across all 24 markets, and across different time scales. This aggregation was necessary in order to achieve results that are statistically highly significant. 
In particular, for shorter time windows, or without aggregating across markets, the observed patterns cannot be detected reliably. This may explain why they have not been reported already a decade ago. \\

Guided by the patterns observed in the graphs in fig. 1, we have performed a nonlinear regression of the next-day normalized log-return $R_i(t+1)$ (\ref{return}) against a polynomial of the current trend strength $\phi_{i,k}(t)$ (\ref{phi}) across all markets $i$ and time scales $T=2^k$:
\begin{equation}
R_i(t+1)=a+b\cdot  \phi_{i,k}(t)+c \cdot \phi_{i,k}^3(t)+\epsilon_i(t+1),\label{cubic}
\end{equation}
where $\epsilon$ represents random noise, and $a$ measures the average normalized risk premium $\mu_i/\sigma_i$ across all assets. These risk premia have not been not the focus of \cite{schmidhuber}. Instead, we have concentrated on determining the coefficients $b, c,$ 
which measure how the expected return of an asset varies in time. We interpret $b$ as the persistence of trends, and $c$ as the strength of trend reversion. 
$d$ and the higher-order terms were not statistically significant. The results of the overall regression analysis are shown in table 2.  \\

\begin{tabular}{ |p{2.5cm}||p{3.5cm}|p{2.5cm}|p{2.5cm}|  }	\hline
 	{\it Table 2} & \multicolumn{3}{|l|}{Regression with linear and cubic terms} \\ \hline
	{Coefficient}& Value  &Error &t-statistics\\	\hline
	$a$   & $1.33$\%    &$\pm 0.41$\% &  $3.3$\\  
	$b$ &$1.29$\% & $\pm 0.43$\% &  $3.0$ \\	
	$c$ &  $-0.62$\%  &  $\pm 0.23$\% & $2.7$\\ \hline\hline
	{R-squared} & Single time scales & \multicolumn{2}{|l|}{Aggregated across time scales} \\ \hline
	$R^2$ &$1.31$\ bp & \multicolumn{2}{|l|}{$4.91$\ bp}  \\	
	$R^2_{adj}$ &  $1.03$\ bp  &  \multicolumn{2}{|l|}{$3.98$\ bp}  \\ \hline
\end{tabular}\\ \\

The coefficients $a,b,c$ and the R-squared's are very small, but they {\it must} be so by nature: if the patterns were much stronger, traders would immediately see and exploit them and achieve stellar Sharpe ratios. In fact, it is shown in \cite{schmidhuber} that 
the estimated coefficients are just strong enough to explain the Sharpe ratios of order 1 that the Managed Futures (``CTA'') industry has historically achieved.\\ 

\cite{schmidhuber} also reports further analyses, in which $b$ and $c$ are allowed to have a simple polynomial dependence on the log of the trend's time horizon  $k\equiv\log_2 T$: 
\begin{equation}
R_i(t+1)=\alpha_i+b(k)\cdot  \phi_{i,k}(t)+c(k) \cdot \phi_{i,k}^3(t)+\epsilon_i(t+1).\label{model}
\end{equation}
Within the limits of statistical significance, we find that $b$ and $c$ are universal, i.e., the same for all assets. 
While the strength of reversion $c$ seems to be approximately constant, we find that the persistence of trends $b$ depends on the time horizon of the trend. 
Within the range of time scales considered here, and averaged over the past 30 years, it can be approximated by a parabolic function of the log of the time scale:
\begin{equation}
c\sim -0.6\%\ \ \ ,\ \ \ b(k)\sim 2.0\%\cdot\Big(1-{(k-k_0)^2\over\Delta k^2}\Big)\ \ \ \text{with}\ \ \ k_0\sim 6\ ,\ \ \ \Delta k\sim 5.\label{para}
\end{equation}
This implies that trends may only be stable if the log of the time horizon is within the range $k_0\pm\Delta k$, corresponding to time scales from a few days to several years. 
The parameters $k_0$ and $\Delta k$ are also universal. By bootstrapping and cross-validation, we have found that all four parameters in  (\ref{para}) are statistically highly significant out-of-sample.

\subsection{Statistical analysis}

Given the small values of the coefficients, and the huge amount of noise in financial market data, we had to pay great attention to a careful analysis of the statistical significance of the results. 
In particular, since market returns cannot be assumed to be independent, identically distributed normal variables, we cannot trust the usual estimates of the t-statistics, adjusted R-squared's, and other statistics. Instead, the test statistics shown in table 2 have been measured empirically as follows:
\begin{itemize}
\item  The standard errors of the coefficients and their t-statistics were computed by bootstrapping, using 5000 bootstrapping samples. Regression on each new sample of days yielded the distribution of the 5000 regression coefficients $b$ and $c$ shown in fig. 1 (upper right), from
which the estimation errors in table 2 were read off. These actual errors are 3-4 times as large as the errors that standard regression tools reported. Nevertheless, the regression coefficients are still statistically highly significant. 
\item The adjusted R-squared was computed by 15-fold cross validation. The resulting out-of-sample R-squared is reported as $R^2_{adj}$ in table 2. 
The actual out-of-sample correction $R^2-R^2_{adj}$ is about 3 times as big as what standard regression tools reported. Nevertheless, the remaining $R^2_{adj}$ is still highly significant despite being very small, and in fact agrees with the one that is achieved in real trading. 
\item  
Table 2 also reports ${R}^2$ and ${R}^2_{adj}$ "aggregated across time scales". Those are based on using the equally-weighted mean of the 10 trend strengths on each day to predict the next-day return for each market. I.e., we combine the 10 different trend factors into a single one, which has a higher predictive power than each single factor by itself. 
\end{itemize}

The regression results of table 2 and further regression analyses in \cite{schmidhuber} confirm and quantify our conclusions from fig. 1. In particular, we see that the values of $b$ and $c$ - although very small - are statistically highly significant, despite the fact that market returns are neither normally distributed, nor independent, nor identically distributed.

\end{document}